\documentstyle[12pt]{article}
\def\bea{\begin{eqnarray}}
\def\eea{\end{eqnarray}}

\def\beq{\begin{equation}}
\def\eeq{\end{equation}}
\def\ba{\beq\new\begin{array}{c}}
\def\ea{\end{array}\eeq}
\def\be{\ba}
\def\ee{\ea}

\makeatletter
\newdimen\normalarrayskip              
\newdimen\minarrayskip                 
\normalarrayskip\baselineskip
\minarrayskip\jot
\newif\ifold             \oldtrue            \def\new{\oldfalse}
\def\arraymode{\ifold\relax\else\displaystyle\fi} 
\def\eqnumphantom{\phantom{(\theequation)}}     
\def\@arrayskip{\ifold\baselineskip\z@\lineskip\z@
     \else
     \baselineskip\minarrayskip\lineskip2\minarrayskip\fi}
\def\@arrayclassz{\ifcase \@lastchclass \@acolampacol \or
\@ampacol \or \or \or \@addamp \or
   \@acolampacol \or \@firstampfalse \@acol \fi
\edef\@preamble{\@preamble
  \ifcase \@chnum
     \hfil$\relax\arraymode\@sharp$\hfil
     \or $\relax\arraymode\@sharp$\hfil
     \or \hfil$\relax\arraymode\@sharp$\fi}}
\def\@array[#1]#2{\setbox\@arstrutbox=\hbox{\vrule
     height\arraystretch \ht\strutbox
     depth\arraystretch \dp\strutbox
     width\z@}\@mkpream{#2}\edef\@preamble{\halign
\noexpand\@halignto
\bgroup \tabskip\z@ \@arstrut \@preamble \tabskip\z@ \cr}%
\let\@startpbox\@@startpbox \let\@endpbox\@@endpbox
  \if #1t\vtop \else \if#1b\vbox \else \vcenter \fi\fi
  \bgroup \let\par\relax
  \let\@sharp##\let\protect\relax
  \@arrayskip\@preamble}
\def\eqnarray{\stepcounter{equation}%
              \let\@currentlabel=\theequation
              \global\@eqnswtrue
              \global\@eqcnt\z@
              \tabskip\@centering
              \let\\=\@eqncr
              $$%
 \halign to \displaywidth\bgroup
    \eqnumphantom\@eqnsel\hskip\@centering
    $\displaystyle \tabskip\z@ {##}$%
    \global\@eqcnt\@ne \hskip 2\arraycolsep
         $\displaystyle\arraymode{##}$\hfil
    \global\@eqcnt\tw@ \hskip 2\arraycolsep
         $\displaystyle\tabskip\z@{##}$\hfil
         \tabskip\@centering
    &{##}\tabskip\z@\cr}

\begin{document}

\begin{titlepage}
\setcounter{footnote}0
\begin{center}

\vspace{0.3in}
{\LARGE\bf Branes and integrability in the N=2 SUSY YM theory}

\bigskip {\Large A.Gorsky
\\
\bigskip
ITEP,Moscow,117259,B.Cheryomushkinskaya 25}
\end{center}
\bigskip

\begin{abstract}
We suggest the brane interpretation of the integrable dynamics behind the
exact solution to the N=2 SUSY YM theory.Degrees of freedom
of the Calogero type integrable system
responsible for the appearance of the spectral Riemann
surfaces originate from  the collective
coordinates of the dynamical branes.  The
second Whitham type integrable system  corresponds to
the low energy scattering of
branes similar to the scattering of the magnetic monopoles.

\end{abstract}

\end{titlepage}

\newpage
\setcounter{footnote}0

1.
Since the beautiful derivation of the low energy effective action for N=2
SUSY YM theories \cite{sw} it was recognized that the full set of the
relevant information about the solution is encoded in the pair of the
integrable systems responsible  for  pure YM theory \cite{Toda1,Toda2},
as well as
YM with adjoint \cite{cal} or fundamental matter \cite{chains}.The solution
was formulated in terms of the moduli spaces relevant for the
vacuum configuration.Therefore the revealing of the integrable systems
can be expected from the very beginning
since they typically have different moduli spaces as the
configuration or  phase spaces.Nevertheless the interpretation of
the field theory data in terms of the integrable systems looks as
the amusing coincidence and the proper identification of the degrees of
freedom in the integrable dynamics has to be found.

In this note we  suggest  interpretation of the integrable systems
in terms of the brane configuration.Namely
in IIA picture one has to involve D0-branes and
D4-branes
which provide the dynamical degrees of freedom
as well as NS5 branes which
represent the background.It turns out that Toda or Calogero  dynamics
comes from the D0 and D4 brane degrees of freedom while the Whitham dynamics
is related to the slow evolution of D4 branes
along the Coulomb branch of the moduli space.We also briefly discuss
the corresponding IIB/F theory picture.

2.
To get the identification of the degrees of freedom  recall the
equation of the vacuum curve for the pure   SU(N) YM theory.
\be
z+\frac{\Lambda}{z}=P_{n}(x) ,
\ee
where $P_{n}$ is  the n-th order polynomial.
Variable $z$ is defined on the spectral curve - n-sheet cover of the
sphere with the pair of the marked points.  The base sphere
can be considered as the degeneration of the torus with a marked point which
is the bare spectral curve for the elliptic Calogero model.
Remind that the modular parameter of the torus is the bare
coupling constant in the N=4 SYM theory.

In IIA picture
perturbatively one deals with the pair of NS5 branes with
worldvolume $(x_{0},...,x_{5})$ and $N_{c}$ D4 branes with worldvolume
$(x_{0},...,x_{3},x_{6})$ stretched between them \cite{Mwit}
(see also \cite{vafa}).The
distance between NS5 branes in $x_{6}$ direction is $\frac{1}{g^{2}}$.
Perturbatively, coordinates of D4 branes in $x_{4}+ix_{5}$ provide
the momenta in the integrable system.However to introduce
the coordinates in the dynamical system additional $N_{c}$ D0 branes
branes located at D4 branes one per each have to be added.These
D0 branes represent nonperturbative effects in the field theory.
In the M theory these degrees of freedom correspond to the KK
excitations on M5 brane wrapped around the spectral curve.

In IIB/F picture we consider  $N_{c}$ parallel
7-branes
wrapped around bare torus (or sphere in the Toda case) resulting in the
SL(N,C) theory on their world wolume  \cite{witbran}.
It was argued \cite{vafatop} that 7-brane coordinates
normal to the torus are twisted so the (1,0) form $\Phi$ on the surface
$\Sigma$
fixes
the spectral curve via equation $det(\Phi(z)-x)=0$.
It can be identified with the Lax operators for the Toda or Calogero
systems. Having in mind the relation to the positions in the normal
direction we can identify its diagonal elements ,playing the role of the
momenta of $N_{c}$ particles $p_{i}$ as the positions of 7-branes
while the nondiagonal elements are due to the strings stretched between the
neighbour 5-branes in the Toda case and between all  in the Calogero case.The
antiholomorphic (0,1) form $\bar{A}$
representing the nonperturbative effects in the field theory is
defined on
$\Sigma$ is diagonal SL(N,C) matrix whose entries $x_{i}$ denote the
coordinates of the particles in the dynamical system .

The Wilson line in the proper representation is
inserted at the marked point on the torus and
appears in the r.h.s of the momentum map equation  \cite{hitchin}
\be
\bar{\partial} \phi +[\bar{A},\phi]=M(1-\delta_{ij})\delta(z).
\ee
This equation concerns the (1,0) form $\Phi$ and (0,1) form
$\bar{A}$ defined on the elliptic curve with the marked point.The
pair $(\Phi,\bar{A})$ with the momentum map restriction provides
the phase space for the elliptic Calogero model.The parameter
M which fixes the monodromy around the marked point corresponds to
the mass of the adjoint hypermultiplet in the 4d field theory.

Resolution of the momentum map equation results in the expression for the
Calogero Lax operator
\be
L_{ij}(Z)=\Phi_{ij}(z)=
\delta_{ij} p_{j}+iM(1-{\delta}_{ij})
exp(\frac{(x_{i}-x_{j})(\lambda-\bar{\lambda})}{\tau-\bar{\tau}})
\frac{\sigma(x_{i}-x_{j}-z)}
{\sigma(z)\sigma(x_{i}-x_{j})} .
\ee
The position of the Wilson line on $\Sigma$ represents the
common coordinate of the $N_{c}$ background branes on the torus and  its
diagonal elements define the coordinates of the
background branes along the direction
of the Wilson line.

3.
Let us find now the interpretation of the equation of motion
in  the Toda case.The D0 branes on D4 branes in IIA picture behave
as monopoles.These D0 branes located on some D4 branes are connected
with the other D4 branes by the open strings.
Following
\cite{diac} we can identify the world volume theory of strings as coming
from 10d SYM theory via  the dimensional reduction.
Moreover there is one to one
correspondence between the equation defining the ground state in the SUSY
$\sigma$
model and the Nahm equations describing the  monopole moduli
space.Finally to get the Toda equation in the brane terms
we use the observation \cite{tfm} that the spectral curve for SL(N,C)
Toda system coincides with the one for the cyclic N-monopoles.
The spectral curve for the cyclic N-monopole follows from the generic
spectral curve
\be
\eta^{n}+\eta^{n-1}a_{1}({\kappa})+....+a_{n}(\kappa)=0
\ee
if one assumes that the center of mass is fixed at the origin,total
phase is unity and the symmetry under the cyclic group of order N is
imposed.

These arguments
result in the following expression
for the Toda Lax operator in terms of the Nahm matrixes $T_{i}$
\be
T=T_{1}+iT_{2}-2iT_{3}{\rho}+(T_{1}-iT_{2}){\rho}^{2}
\ee
\be
T_{1}=\frac{i}{2}\sum_{j=1}q_{j}(E_{+j}+E_{-j})  \\
T_{2}=-\sum_{j=1}q_{j}(E_{+j}-E_{-j})             \\
T_{3}=\frac{i}{2}\sum_{j}p_{j}H_{j} ,
\ee
where E and H are the standard SU(N) generators, $p_{i},q_{i}$ represent
the Toda phase space,and $\rho$ is the coordinate on the $CP^{1}$ above.
This $CP^{1}$ is involved in the twistor construction for
monopoles and a point on $CP^{1}$  defines the complex structure on the
monopole moduli space.
With these definitions Toda equation of motion and Nahm
equation acquire the simple form
\be
\frac{dT}{dt}=[T,A]
\ee
with the fixed A. Note that to define A operator it is necessary to
introduce
the fourth Nahm operator $T_{0}$.

To get the additional insight on the nature of the variables
it is useful to consider the fermions in the external field.
Namely let us consider 3d
Dirac operator in the external field
\be
({\sigma}D+{\phi}-t){\Psi}=0 ,
\ee
which has k linearly independent solutions for the magnetic charge k.That is
just the variable t becomes the time variable in the Toda dynamics.The Nahm
matrixes come from the fermions in a wellknown way
\be
T_{j}=\int x_{j}{\Psi}^{+}{\Psi} d^{3}x;j=1,2,3     \\
T_{4}=i\int{\Psi}^{+}\frac{d{\Psi}}{dt}d^{3}x  .
\ee
Both expressions have the form of the Berry connection and the
momentum space serves as the space of parameters.In the inverse construction
one can derive the world sheet monopole fields starting with the
auxiliary fermions in the background  Nahm connection.The proceedure
looks as follows;
\be
({i\frac{d}{dt}} +i\sigma x+\sigma T^{+})v=0
\ee
\be
A_{j}=\int v^{+}{\partial}_{j}vdt \\
\phi=\int tv^{+}vdt ,
\ee
where the integration intervals are defined through the asymptotic values
of the Higgs field.Fermions
taken at x=0 look like the fermions in the auxiliary spectral problem for the
Toda system.Therefore the analogy with the Peierls model discussed in
\cite{pei} can be partially clarified.Namely equation for the fermions
in the Nahm connection can be treated as the Schr${\ddot o}$dinger equation
while the 1d "crystal" with $N_{c}$ cites
is formed by the D0 branes.
Cyclic symmetry of the monopole configuration implies the periodicity of the
system so the quasimomentum can be introduced for the fermions.
Spectral curve
of the Toda system becomes the dispersion law for the fermions and
the field theory BPS state gets interpreted as the completely filled
band.
Integrals of motion in the Toda system parametrize the moduli
space of the strongly centered cyclic monopoles and can be expressed in terms
of the vacuum expectation values of the scalar field in the YM theory.

From the brane picture we get some
insight on the second Lax representation for the Toda system by
2${\times}$2 matrixes.Remind that Toda transfer matrix for the $ SU(N_{c})$
case has the following form
\be
T=\prod_{i=1}^{N_{c}} L_{i}(\lambda-\lambda_{i})
\ee
$$
L_{i}=\left (
\begin{array}{cc}
p_{i}-\lambda & exp(\phi_{i})\\
exp(-\phi_{i}) & 0
\end{array}
\right )
$$
therefore the momenta define the positions of  D0 branes
in $x_{4}+ix_{5}$ direction
while nondiagonal elements are in the correspondence with the
instanton-like fermion
transitions between the nearest sites.This picture is suggestive for the
generalization to the theory with the fundamental matter governed by the spin
chains \cite{chains}.Indeed for this case one keeps the size of the local L
operator assuming the elements of the matrix to be SL(2,R) valued.

4.
To recognize the second Whitham type integrable system  let us adopt
slightly different perspective
from the F-theory on the elliptically fibered K3 which is equivalent
to the orientifold of type IIB theory or, after T duality, to type I theory
on $T^{2}$.
Due to \cite{sen} we can treate the N=2 d=4 theory as a world volume theory
of probe
3-branes in the background of the splitted orientifold 7 planes placed at
points $\pm\Lambda$ in the $u=Tr\phi^{2}$ complex plane
for the SU(2) case.We assume that the possible masses
of the fundamental matter tend to infinity so we work with the pure YM case.
Note that actually we have $SL(2,C)$ bundle on the elliptic fiber in
the F theory,which is the torus with a marked point.Degeneration
of the fiber to the sphere corresponding to the pure YM theory
is performed in a way providing the emergence of the dimensional
transmutation parameter $\Lambda$.

The key point is that now we have to consider the dynamics of the
3-branes
in the directions transverse to the background 7 branes.
The arising dynamics is quite
transparent already in the SU(2) case.Let us recall that Whitham dynamics
for SU(2) case is governed by the solution of the first Gurevich-Pitaevskii
problem \cite{Toda1}
which can be easily interpreted as follows.At the initial moment of
evolution 3-branes coincide with the one of the orientifold planes and with the
other planes at the end of the evolution.Therefore the exact metric on the
moduli space can be derived from the 3-brane exchange between the orientifold
planes which extends the closed string exchange picture  \cite{doug}
and gives rise to the perturbative part of the metric on the moduli space.
To analyze the Whitham dynamics in SU(2) case it is convenient to use the
following form of the spectral curve
\be
y^{2}=(x^{2}-\Lambda^{2})(x-u).
\ee
The point u represents the position of two 3-branes (which are at
$\pm \sqrt u$
at  $\phi$ plane) and "nonperturbative" branching points  provide the fixed
positions of the background branes.The branching point u
and therefore a pair of 3-branes moves
from $ u=\Lambda$ to $ u=-\Lambda$ according to
the Whitham dynamics for the one-gap KdV solution which
corresponds to the Seiberg-Witten solution.  \be
\eta(x,t)=2dn^{2}[\frac{1}{\sqrt6}(x-\frac{1+s^{2}}{3}t,s)]-(1-s^{2})
\ee
where
\be
\frac{1+s^{2}}{3}
-\frac{2s^{2}(1-s^{2})K(s)}{3(E(s)-(1-s^{2})K(s))}=\frac{x}{t},
\ee
K(s) and E(s) are the elliptic moduli and
$s^{2}=\frac{u+\Lambda}{2\Lambda}$.
In terms of the automodel variable $\theta=\frac{x}{t}$
the left background brane corresponds to $\theta=-1$ while the right one to
$\theta=\frac{2}{3}$.Quasiclassical tau-function of this solution provides
the prepotential for the SU(2) theory $\cal{F}$=log${\tau}_{qcl}$
\cite{Toda1}.

Let us emphasize the close analogy of the motion above
with the monopole scattering in the low energy
limit.It is known that there are several types of the geodesic motion on
the monopole moduli space which governs the low-energy monopole scattering
\cite{book}.
Among them there is a  geodesic motion which results in the
transformation
of the initial monopoles into the dyons.Let us compare it with the
Whitham dynamics under consideration.
In the $\phi$ plane there are two dynamical 3-branes placed at $\pm\phi$
and four background branes placed at $\pm \sqrt \Lambda;\pm i\sqrt \Lambda$.
At the initial moment one has the massless monopoles
in the spectrum since the dynamical branes are placed at
$\pm \sqrt{\Lambda}$.
After
the right angle scattering at u=0 3-branes acquire the electric charges.
Indeed we know that at the final point of the Whitham evolution
$u=-\Lambda$ one has the massless dyons which can be considered
as the (1,1) string stretched between the 3-brane and  background brane.
Therefore the string can end on the 3-brane only if it has  the electric
charge.

5.
To conclude we have shown that degrees of freedom in the finite-dimensional
integrable systems relevant for SYM theories allow the interpretation
in  terms of branes.Collective coordinates of branes cover the phase space
of the integrable system so the question if the
quantization of  the integrable system is meaningful gets the positive
answer.It would correspond to the collective coordinate quantization
which provides the proper wave function.Moreover the interpretation
of the Whitham dynamics in terms of the brane scattering implies that
a kind of "quantization" can provide a
description of the "quantum scattering".
It would be also important to recognize the direct relation between the
monopole moduli spaces considered in this note and the ones appeared
in the treatment of the SUSY 3d theories \cite{3d}.One more interesting
issue to be discussed is
the brane interpretation of the relativistic integrable system found
in 5d theory \cite{5d}.

I thank to TPI at   University of Minnesota where this paper was
completed for the kind hospitality.I am indebted to
N.Nekrasov,A.Vainshtein and
M.Voloshin for the useful discussions.  The work was partially supported by
the grant INTAS 93--2494.

\end{document}